\documentclass[journal]{IEEEtran}

\usepackage{cite}
\usepackage{url}

\usepackage{graphicx}
\usepackage[cmex10]{amsmath}
\usepackage{amsfonts}

\usepackage{array}
\usepackage{subfig}
\usepackage{booktabs}
\usepackage{here}

\makeatletter
\def\Hline{%
\noalign{\ifnum0=`}\fi\hrule \@height 2pt \futurelet
\reserved@a\@xhline}
\makeatother

\def\Vec#1{\mbox{\boldmath $#1$}}
\def\figref#1{Fig.~\ref{#1}}

\def\tabref#1{Table~\ref{#1}}

\def\eqref#1{(\ref{#1})}
\def\Eqref#1{Eq.~\ref{#1}}

\newcommand{\argmin}{\mathop{\rm arg~min}\limits}

\begin{document}

\title{Joint optimization of multispectral filter arrays and demosaicking for pathological images}

\author{\authorblockN{
Kazuma Shinoda\authorrefmark{1}, Maru Kawase\authorrefmark{1},  Madoka Hasegawa\authorrefmark{1}, 
Masahiro Ishikawa\authorrefmark{2},
Hideki Komagata\authorrefmark{2},
Naoki Kobayashi\authorrefmark{2}} \\
\authorblockA{\authorrefmark{1}Graduate School of Engineering,
 Utsunomiya University, Utsunomiya, Japan} \\
\authorblockA{\authorrefmark{2}Faculty of Health and Medical Care, Saitama Medical University, Hidaka, Japan}
}

\markboth{PREPRINT VERSION. IIEEJ Transactions on Image Electronics and Visual Computing, Vol. 6, No. 1, pp. 13-21, Jun. 2018.}{PREPRINT VERSION. IIEEJ Transactions on Image Electronics and Visual Computing, Vol. 6, No. 1, pp. 13-21, Jun. 2018.}

\maketitle

\begin{abstract}
A capturing system with multispectral filter array (MSFA) technology is proposed for shortening the capture time and reducing costs. Therein, a mosaicked image captured using an MSFA is demosaicked to reconstruct multispectral images (MSIs). Joint optimization of the spectral sensitivity of the MSFAs and demosaicking is considered, and pathology-specific multispectral imaging is proposed. This optimizes the MSFA and the demosaicking matrix by minimizing the reconstruction error between the training data of a hematoxylin and eosin-stained pathological tissue and a demosaicked MSI using a cost function. Initially, the spectral sensitivity of the filter array is set randomly and the mosaicked image is obtained from the training data. Subsequently, a reconstructed image is obtained using Wiener estimation. To minimize the reconstruction error, the spectral sensitivity of the filter array and the Wiener estimation matrix are optimized iteratively through an interior-point approach. The effectiveness of the proposed MSFA and demosaicking is demonstrated by comparing the recovered spectrum and RGB image with those obtained using a conventional method.
\end{abstract}



\section{Introduction} \label{sec:intro}

A pathologist is a medical doctor specializing in the diagnosis and classification of diseases by examining tissue or cells under a microscope.
Because human tissues are relatively transparent unless they contain an endogenous pigment, the pathological tissues are generally stained using hematoxylin and eosin (H\&E).
In computer-aided diagnosis, the stained tissue is captured using a digital camera under a microscope, the resulting image of which is called a ``pathological image,'' and various tissues such as the nucleus (stained blue) and cytoplasm and fiber (stained red) can be observed from a pathological image.
In recent years, multispectral images (MSIs) of pathological tissues have been studied.
Abe proposed a color-correction method for H\&E-stained pathological images using a 16-band multispectral microscope camera system \cite{ref:TAbe2005}. Color reproduction and morphological characteristics of the nuclei/cells are important in pathological diagnosis. Another study \cite{ref:MTashiro2009} examined the color differences in the nuclei using a multispectral imaging system. Because the spectral features of pathological tissues can be estimated from MSIs, some studies have explored the digital staining of pathological tissues \cite{ref:PABautista2015}.

However, limitations persist in the techniques used for capturing MSIs owing to the complexity of assembling prisms or multiple sensor arrays for signal detection. Because most current multispectral cameras require a relatively large amount of time per frame (a few seconds or more), reducing the capturing time while retaining the image quality is an important factor in the popularization of pathological imaging systems. Considering the application of color filter arrays (CFAs) in commercial digital RGB cameras, multispectral filter arrays (MSFAs) have been studied to resolve this issue, and various filters and demosaicking methods have thus been proposed \cite{ref:JBrauers2006,ref:YMonno2015,ref:FYasuma2010,ref:HKAggarwal2013}.

\begin{figure}[!t]
 \centerline{
  \subfloat[]{\includegraphics[width = 0.35\linewidth]{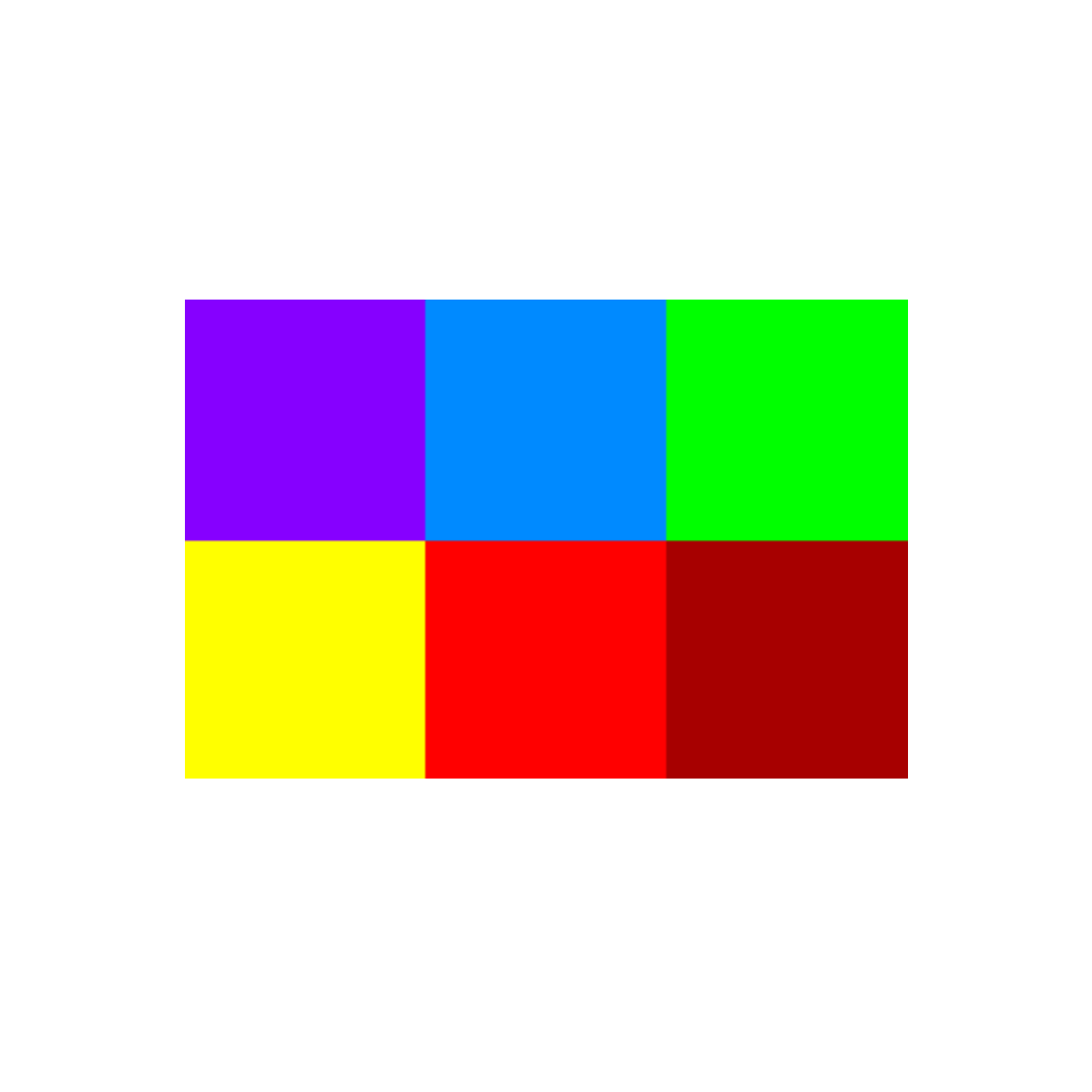}%
  \label{fig:BrauersMSFA}}
  \hfil
  \subfloat[]{\includegraphics[width = 0.35\linewidth]{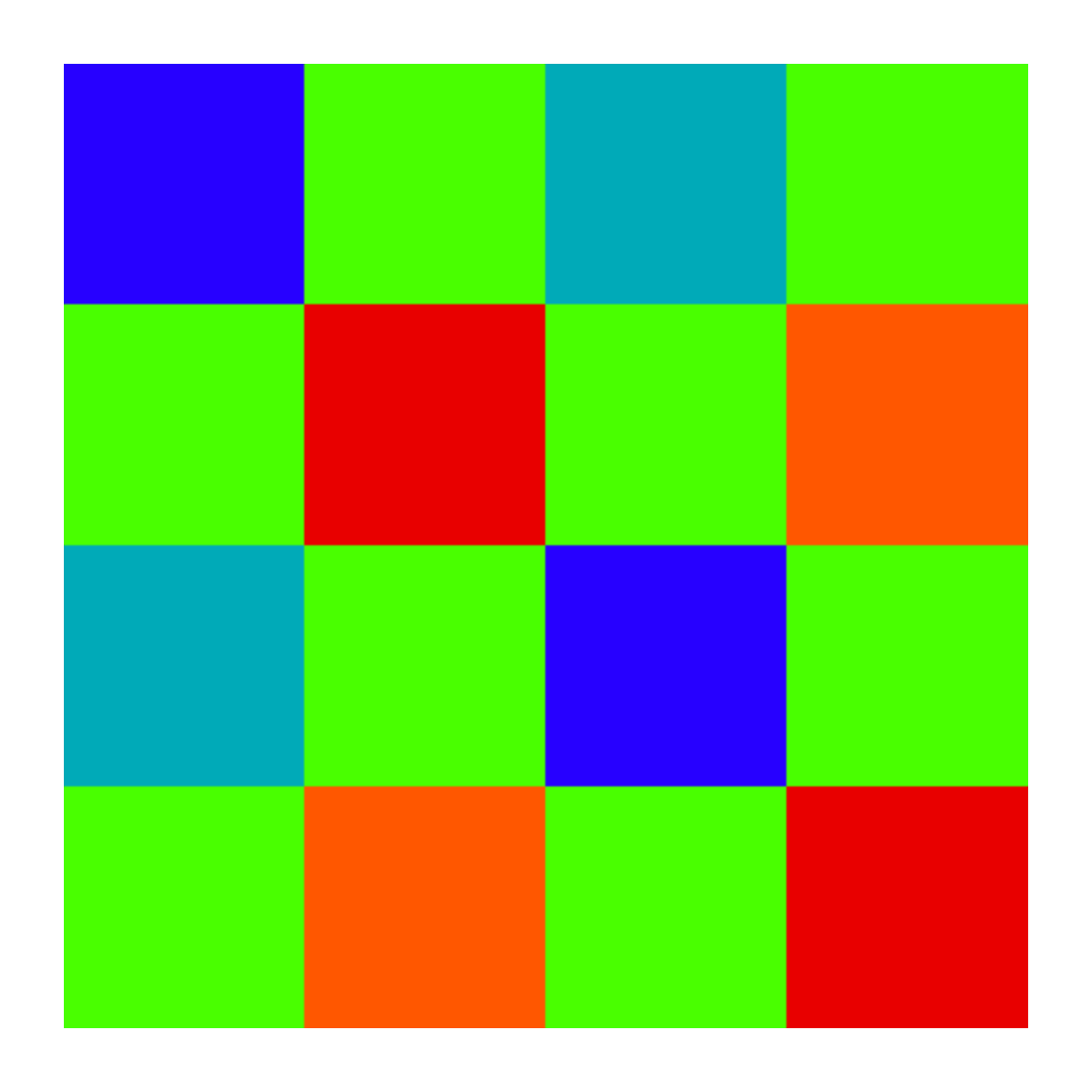}%
  \label{fig:MonnoMSFA}}
 }
 \caption{Examples of MSFAs (a) Brauers and Aach \cite{ref:JBrauers2006}, (b) Monno et al. \cite{ref:YMonno2015}}
 \label{fig:ExamplesOfMSFA}
\end{figure}

\begin{figure*}[!t]
 \begin{center}
  \includegraphics[width = 0.8\linewidth]{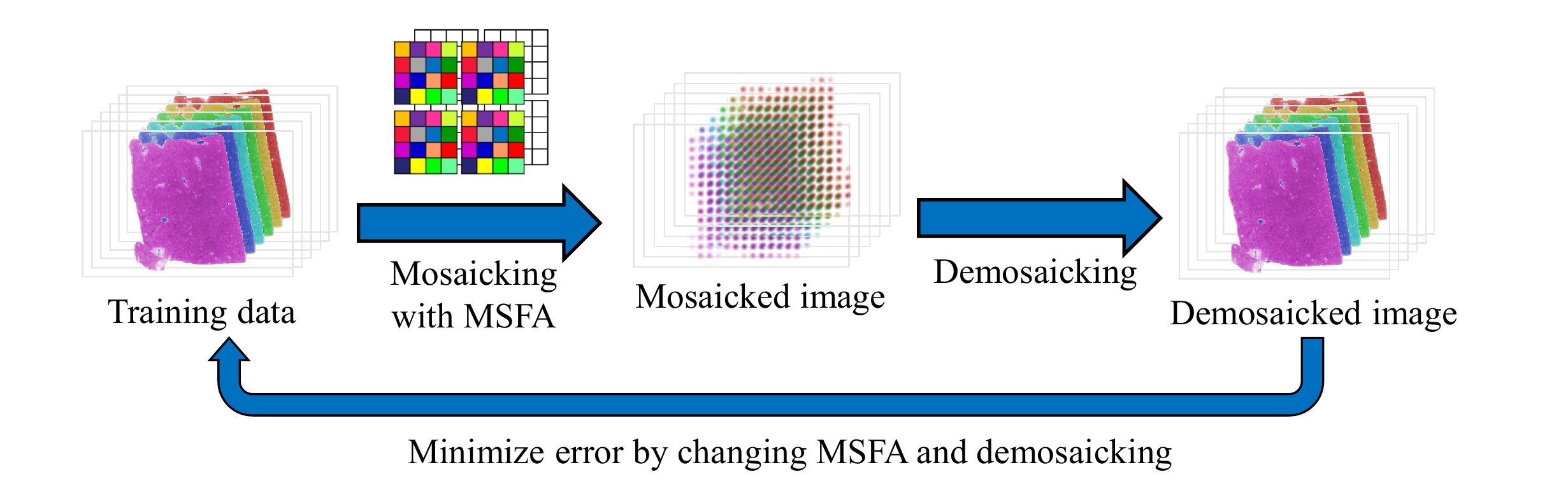}
 \end{center}
 \caption{Designing MSFA and demosaicking using full-resolution MSI}
 \label{fig:MSFAsimulation}
\end{figure*}

Two example MSFAs are shown in \figref{fig:ExamplesOfMSFA}. Brauers et al. \cite{ref:JBrauers2006} proposed a six-band MSFA arranged with a pixel resolution of $3\times2$ in a straightforward manner intended for faster linear interpolation. Monno et al. \cite{ref:YMonno2015} proposed a five-band MSFA, and determined that the sampling density of the G-band data was higher than that of the other spectral bands because the human eye is more sensitive to the G-band. However, they did not consider specific applications such as pathological images. The color distribution of pathological images is biased toward blue and magenta because these colors consist of a limited number of dyes and the variety of spectral distributions of stained cells is also considerably limited. Therefore, the quality of a demosaicked image can be improved by optimizing the MSFA design and the demosaicking method for pathological applications.

When full-resolution MSIs, which are not mosaicked, are employed, the MSFA pattern and the demosaicking matrix can be optimized by minimizing the error between the original and demosaicked images during a simulation, as shown in \figref{fig:MSFAsimulation}. This approach has been explored in remote sensing applications to jointly capture RGB and near-infrared (NIR) images. Lu et al. \cite{ref:YMLu2009} formulated an MSFA as an optimization problem in the spatial domain, and provided an iterative procedure to locally obtain the optimal solutions. The resulting mosaic pattern consisted of 16 filters arranged in a $4\times4$ pixel resolution, 15 of which were visible and one was in the NIR. An improved algorithm was later developed that considers the correlation between the visible and NIR bands \cite{ref:ZSadeghipoor2011}. Monno et al. \cite{ref:YMonno2015} also developed optimal spectral sensitivity functions of five-band MSFAs. Their method iteratively optimizes the parameters of the Gaussian distribution of each filter. Our previous study \cite{ref:KShinoda2015,ref:YYanagi2016} assumed that the spectral sensitivity of each filter was an ideal delta function, and simultaneously optimized the number of filters and the center wavelength of each one. However, the optimal spectral sensitivity of each filter with respect to a specific application was not considered. If the spectral sensitivity and demosaicking are optimized using pathological images, the demosaicked quality will be improved, and the optimized MSFA and demosaicking may facilitate whole-slide imaging and tele-pathology.

In this study, a joint optimizing method for the spectral sensitivity of an MSFA and the demosaicking for H\&E-stained pathological images is proposed, and the optimized MSFA pattern, demosaicking matrix, and demosaicked quality are exhibited through test images. H\&E-stained human liver tissue was used as the training sample, and a 31-band MSI captured using a line-scan hyperspectral camera was used as an original image. The proposed method first assumes that the mosaicking process using an MSFA corresponds to a linear system, and that the demosaicking process is regarded as an inverse problem of the linear model. The original image is mosaicked using a random MSFA through a simulation; subsequently, the mosaicked image is recovered using a Wiener estimation matrix. The spectral sensitivities of the MSFA are optimized iteratively using an interior-point approach to minimize the errors between the original and demosaicked MSIs. The Wiener estimation matrix is also optimized according to the changes in the spectral sensitivity. After optimizing the MSFA, it is finally demonstrated that the proposed MSFA pattern outperforms a conventional method.

The remainder of this paper is organized as follows: In Section 2, the method used for designing the MSFA and the demosaicking is presented. In Section 3, the experiment results are discussed. Section 4 provides some concluding remarks regarding the proposed method.


\begin{figure}[!t]
 \begin{center}
  \includegraphics[width = 0.8\linewidth]{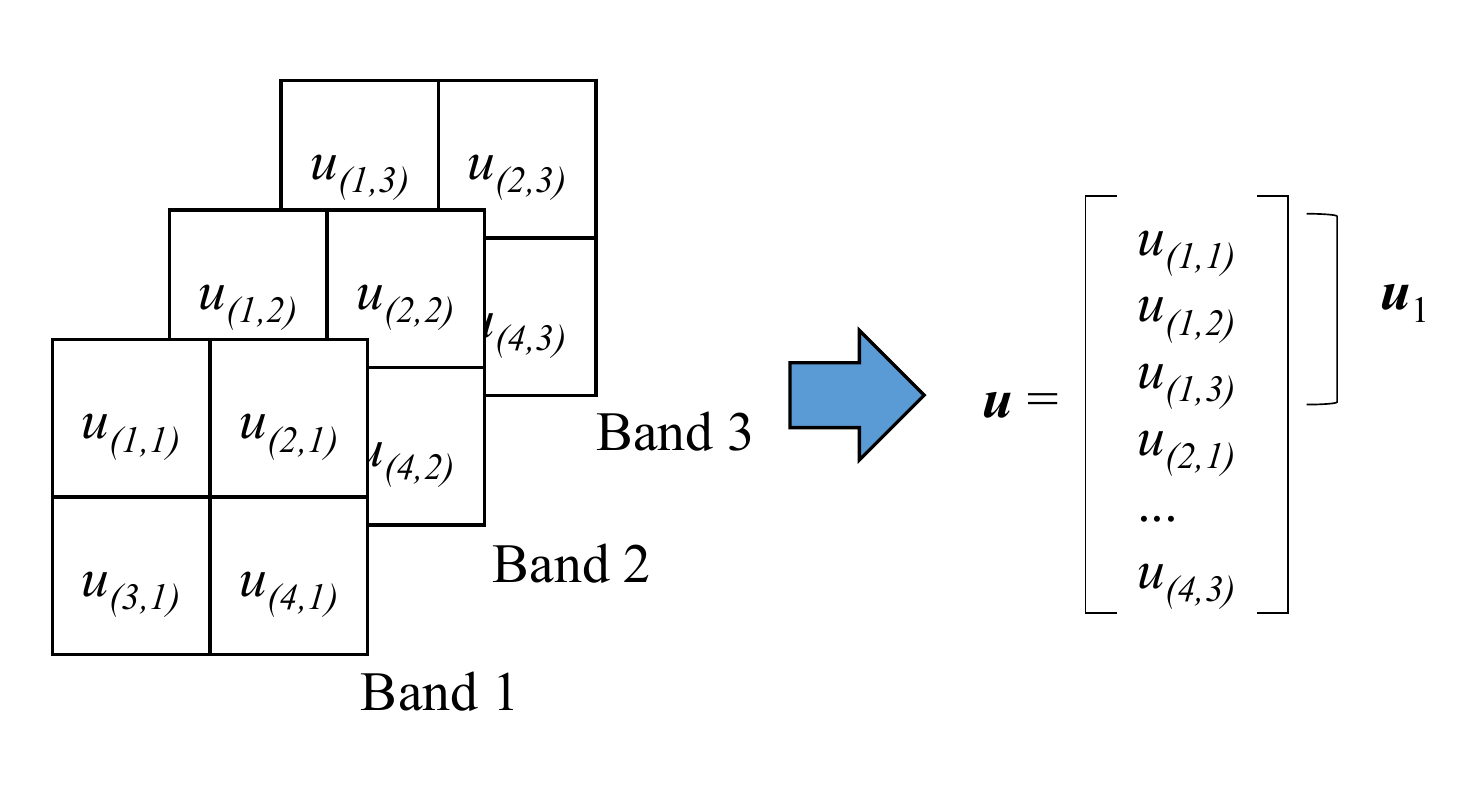}
 \end{center}
 \caption{Example of vectorized pixel values of $\Vec{u}$ ($N = 4$ and $L = 3$)}
 \label{fig:Vector}
\end{figure}

\section{Joint optimization of spectral sensitivity and demosaicking for multispectral filter arrays}
The model for measuring a mosaicked MSI using an MSFA and the recovery problem are first formulated.
We assume that a small MSFA block is arranged periodically for covering the entire image.
The original vectorized MSI $\Vec{u} \in \mathbb{R}^{LN}$, which is not mosaicked, is given as follows (in the remainder of this paper, all vectors are defined as a column vector):
\begin{eqnarray}
\Vec{u} &:=& \left[ \Vec{u}_{1}^{T} \ \Vec{u}_{2}^{T} \ldots \  \Vec{u}_{N}^{T} \right]^{T} \\
\Vec{u}_{n} &:=& \left[ u_{(n,1)} \ \ u_{(n,2)} \ \ \ldots \ \ u_{(n,L)} \right]^{T},
\end{eqnarray}
where $N$ is the number of pixels of $\Vec{u}$ within one MSFA block, $L$ is the number of bands, and $^{T}$ indicates the transpose operator. In addition, $u_{(n,l)}$ is the pixel value at the $n$-th spatial position of band $l$.
A detailed example of vectorized pixels is shown in \figref{fig:Vector}.
The measurement model for the mosaicked MSI $\Vec{v} \in \mathbb{R}^{N}$ in one MSFA block is
\begin{eqnarray}
\Vec{v} &=& \Vec{\Phi}\Vec{u}, \label{eq:Mosaicking}
\end{eqnarray}
where $\Vec{\Phi} \in \mathbb{R}^{N\times LN}$ is the measurement matrix corresponding to an MSFA as follows:
\begin{eqnarray}
\Vec{\Phi} &:=& \left[
  \begin{array}{cccc}
\Vec{\phi}_{1}^{T} & & & \Vec{0} \\
 & \Vec{\phi}_{2}^{T} & & \\
 & & \ddots &  \\
 \Vec{0} & & & \Vec{\phi}_{N}^{T}
  \end{array}
\right]. \label{eq:Phi}
\end{eqnarray}
Here, $\Vec{\phi}_{n} \in \mathbb{R}^{L}$ is the spectral sensitivity of an MSFA at the $n$-th spatial position.
The demosaicking of the MSI can be formulated as an inverse problem of \eqref{eq:Mosaicking}:
\begin{eqnarray}
\hat{\Vec{u}} &=& \Vec{W}\Vec{v}, \label{eq:Demosaicking}
\end{eqnarray}
where $\hat{\Vec{u}} \in \mathbb{R}^{LN}$ is the demosaicked MSI. We assume that $\Vec{W}$ can be determined through a Wiener estimation matrix \cite{ref:WKPratt1978} $\Vec{W} \in \mathbb{R}^{LN \times N}$.
Wiener estimation is applied to minimize the mean-square restoration error, and it takes into account the correlation of $\Vec{u}$ as prior information, i.e., 
\begin{eqnarray}
\Vec{W} &=& \Vec{R}_{u}\Vec{\Phi}^{T}\left(\Vec{\Phi}\Vec{R}_{u}\Vec{\Phi}^{T}\right)^{-1}, \label{eq:Wiener}
\end{eqnarray}
where $\Vec{R}_{u} \in \mathbb{R}^{(LN \times LN)}$ is the autocorrelation matrix of $\Vec{u}$.
The joint optimization problem of the MSFA and demosaicking is formulated using \eqref{eq:Mosaicking} and \eqref{eq:Demosaicking} as
\begin{eqnarray}
(\Vec{\Phi}, \Vec{W}) &=& \argmin_{\Vec{\Phi},\Vec{W}} ||\Vec{u} - \Vec{W}\Vec{\Phi}\Vec{u}||_2. \label{eq:Problem}
\end{eqnarray}

To solve \eqref{eq:Problem} efficiently, we first modify the Wiener estimation.
\Eqref{eq:Problem} uses only one block of a mosaicked image $\Vec{\Phi}\Vec{u}$ for estimating $\hat{\Vec{u}}$; however, the estimation error can be reduced by considering the mosaicked pixels in the neighboring blocks.
In brief, when estimating an unknown pixel value, the estimation error will be reduced as the number of neighboring known pixels is increased. Although an excessive spread in the number of pixels used will cause a greater estimation error, spreading the area from one block to $3 \times 3$ blocks has been confirmed to have an advantage in RGB demosaicking \cite{ref:YMLu2009_2}.
Therefore, we extend the mosaicking process of \eqref{eq:Mosaicking} to a process using $3 \times 3$ blocks as follows:
\begin{eqnarray}
\Vec{v}' &=& \Vec{\Phi}'\Vec{u}' \label{eq:Mosaicking2} \\
\Vec{v}' &:=& [\Vec{v}_{(x-1, y-1)}^{T}, \Vec{v}_{(x, y-1)}^{T}, \Vec{v}_{(x+1, y-1)}^{T}, \Vec{v}_{(x-1, y)}^{T}, \nonumber \\ 
 && \ldots, \Vec{v}_{(x, y)}^{T}, \ldots, \Vec{v}_{(x+1, y+1)}^{T}]^{T} \\
\Vec{u}' &:=& [\Vec{u}_{(x-1, y-1)}^{T}, \Vec{u}_{(x, y-1)}^{T}, \Vec{u}_{(x+1, y-1)}^{T}, \Vec{u}_{(x-1, y)}^{T}, \nonumber \\ 
 && \ldots, \Vec{u}_{(x, y)}^{T}, \ldots, \Vec{u}_{(x+1, y+1)}^{T}]^{T},
\end{eqnarray}
where $\Vec{\Phi}' = \Vec{I}_{9} \otimes \Vec{\Phi}$, the index $(x, y)$ is a block-wise index, $\otimes$ is the Kronecker product, and $\Vec{I}_{9}$ is an $9 \times 9$ identity matrix.

\Eqref{eq:Mosaicking2} is extended from the one-block process of \eqref{eq:Mosaicking} to a nine-block process. Therefore, the $\Vec{v}'$ obtained is identical to the concatenation of the nine $\Vec{v}$ vectors of \eqref{eq:Mosaicking}.
The Wiener estimation of \eqref{eq:Mosaicking2} is 
\begin{eqnarray}
\hat{\Vec{u}}' &=& \Vec{W}'\Vec{v}' \\ \label{eq:Demosaicking2}
\Vec{W}' &=& \Vec{R}'_{u}\Vec{\Phi}'^{T}\left(\Vec{\Phi}'\Vec{R}'_{u}\Vec{\Phi}'^{T}\right)^{-1}. \label{eq:Wiener2}
\end{eqnarray}
It should be noted that $\Vec{R}'_{u} \in \mathbb{R}^{(9LN \times 9LN)}$ is the autocorrelation matrix of $\Vec{u}$ ``in nine blocks''.
That is, a demosaicked pixel of $\hat{\Vec{u}}'$ is calculated by utilizing the spatial and spectral correlation in nine blocks.
Therefore, the demosaicked error in the center block of $\hat{\Vec{u}}'$ can be expected to be further reduced compared with \eqref{eq:Demosaicking}.
Because the demosaicked error of $\hat{\Vec{u}}'$ in the outer eight blocks is larger than that in the center block, we extract only the center demosaicked block as follows:
\begin{eqnarray}
\hat{\Vec{u}}_{(x, y)} &=& \Vec{S}\Vec{W}'\Vec{v}', \label{eq:Demosaicking3}
\end{eqnarray}
where $\Vec{S} = [\Vec{0}_{4LN} \Vec{I}_{LN} \Vec{0}_{4LN}]$ is the extraction matrix, and $\hat{\Vec{u}}_{(x, y)}$ corresponds to $\hat{\Vec{u}}$ in \eqref{eq:Demosaicking}.
Therefore, the joint optimization problem of the MSFA and demosaicking can be rewritten as
\begin{eqnarray}
(\Vec{\Phi}', \Vec{W}') &=& \argmin_{\Vec{\Phi}',\Vec{W}'} ||\Vec{u} - \Vec{S}\Vec{W}'\Vec{\Phi}'\Vec{u}'||_2. \label{eq:Problem2}
\end{eqnarray}

\begin{table}[t]
  \begin{tabular}{l} \Hline
    Algorithm 1: Proposed joint optimization solver for spectral \\
     sensitivity $\Vec{\Phi}$ and demosaicking $\Vec{W}'$ \\ \Hline
    {\bf Input}: Training data: $\Vec{u}$ \\
    {\bf Output}: One MSFA block $\Vec{\Phi}$, Wiener estimation matrix \\
     for demosaicking $\Vec{W}'$\\
    {\bf Initialization}: \\
    $\Vec{R}'_{u} \leftarrow$ autocorrelation matrix of  $\Vec{u}$ in nine blocks\\
    $\Vec{\Phi}_{0} \leftarrow$ random values in $[0, 1]$ \\
    {\bf for} $i = 1, 2, \ldots, i_{e}$ \\
    \hspace{10pt} $\Vec{W}'_{i} \leftarrow \Vec{R}'_{u}\Vec{\Phi}'^{T}_{i-1}\left(\Vec{\Phi}'_{i-1}\Vec{R}'_{u}\Vec{\Phi}'^{T}_{i-1}\right)^{-1}$ \\
    \hspace{10pt} $\Vec{\Phi}_{i} \leftarrow \argmin_{\Vec{\Phi}} ||\Vec{u} - \Vec{S}\Vec{W}'_{i}\Vec{\Phi}'_{i-1}\Vec{u}'||_2$ \\
    {\bf end} \\
    $\Vec{\Phi} \leftarrow \Vec{\Phi}_{i_{e}}$ \\
    $\Vec{W}' \leftarrow \Vec{R}'_{u}\Vec{\Phi}'^{T}_{i_{e}}\left(\Vec{\Phi}'_{i_{e}}\Vec{R}'_{u}\Vec{\Phi}'^{T}_{i_{e}}\right)^{-1}$ \\ \Hline
  \end{tabular}
\end{table}

Because $\Vec{W}'$ can be expanded using $\Vec{\Phi}$, the optimization problem can be formulated using only one variable $\Vec{\Phi}$ as follows:
\begin{eqnarray}
&& \Vec{\Phi} = \nonumber \\
&& \argmin_{\Vec{\Phi}} ||\Vec{u} - \Vec{S} \Vec{R}'_{u}(\Vec{I}_{9} \otimes \Vec{\Phi})^{T}\left((\Vec{I}_{9} \otimes \Vec{\Phi})\Vec{R}'_{u}(\Vec{I}_{9} \otimes \Vec{\Phi})^{T}\right)^{-1}  \nonumber \\
&& \ \ (\Vec{I}_{9} \otimes \Vec{\Phi})\Vec{u}'||_2. \label{eq:Problem3}
\end{eqnarray}
However, the non-linear function of \eqref{eq:Problem3} is complicated, and it is therefore difficult to obtain the global minimum value through an exhaustive search. Therefore, the problem in \eqref{eq:Problem2} is solved by considering two separate processes, namely, mosaicking and demosaicking, and optimizing $\Vec{\Phi}$ and $\Vec{W}'$ iteratively, as shown in Algorithm 1.

Initially, multispectral pathological images are given as the training data $\Vec{u}$.
In addition, $\Vec{R}'_{u}$ is calculated for each $3 \times 3$ block from the training data, and $\Vec{\Phi}_{0}$ is initialized with random values in $[0, 1]$.
Note that the invalid area ($\Vec{0}$ vectors) of $\Vec{\Phi}$, shown in \eqref{eq:Phi}, is not changed in the following process.
In the loop, $\Vec{W}'_{i}$ is first calculated based on \eqref{eq:Wiener2}.
Then, $\Vec{\Phi}_{i}$ is optimized under a given $\Vec{W}'_{i}$ based on \eqref{eq:Problem2}.
Although we need the constraint $\Vec{\phi}_{n} \in [0,1]$, the optimization problem of \eqref{eq:Problem2} is a simple quadratic programing problem if $\Vec{W}'_{i}$ is fixed. We used an interior-point approach \cite{ref:RHByrd2000,ref:RAWaltz2006} because doing so can solve the problem effectively within a sufficiently short period of time (less than 1 s in our experiment). At the end of the loop, we can obtain the optimized MSFA $\Vec{\Phi}$ and the Wiener estimation matrix $\Vec{W}'$.

\section{Results and discussion}

\begin{figure}[!t]
 \begin{center}
  \includegraphics[width = 0.9\linewidth]{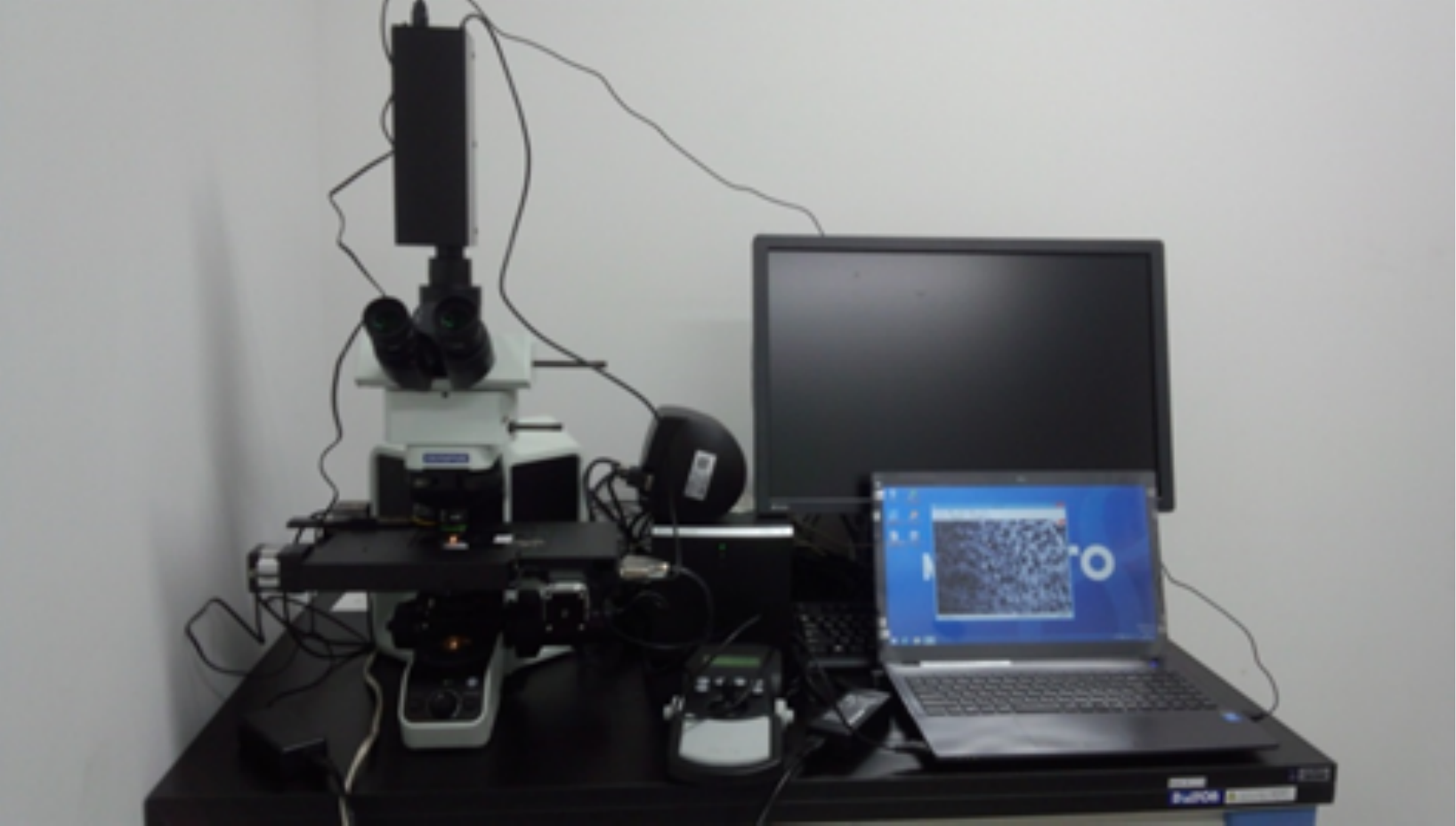}
 \end{center}
 \caption{Hyperspectral imaging system for capturing pathological tissues}
 \label{fig:Setup}
\end{figure}

\begin{figure}[!t]
 \centerline{
  \subfloat[]{\includegraphics[width = 0.7\linewidth]{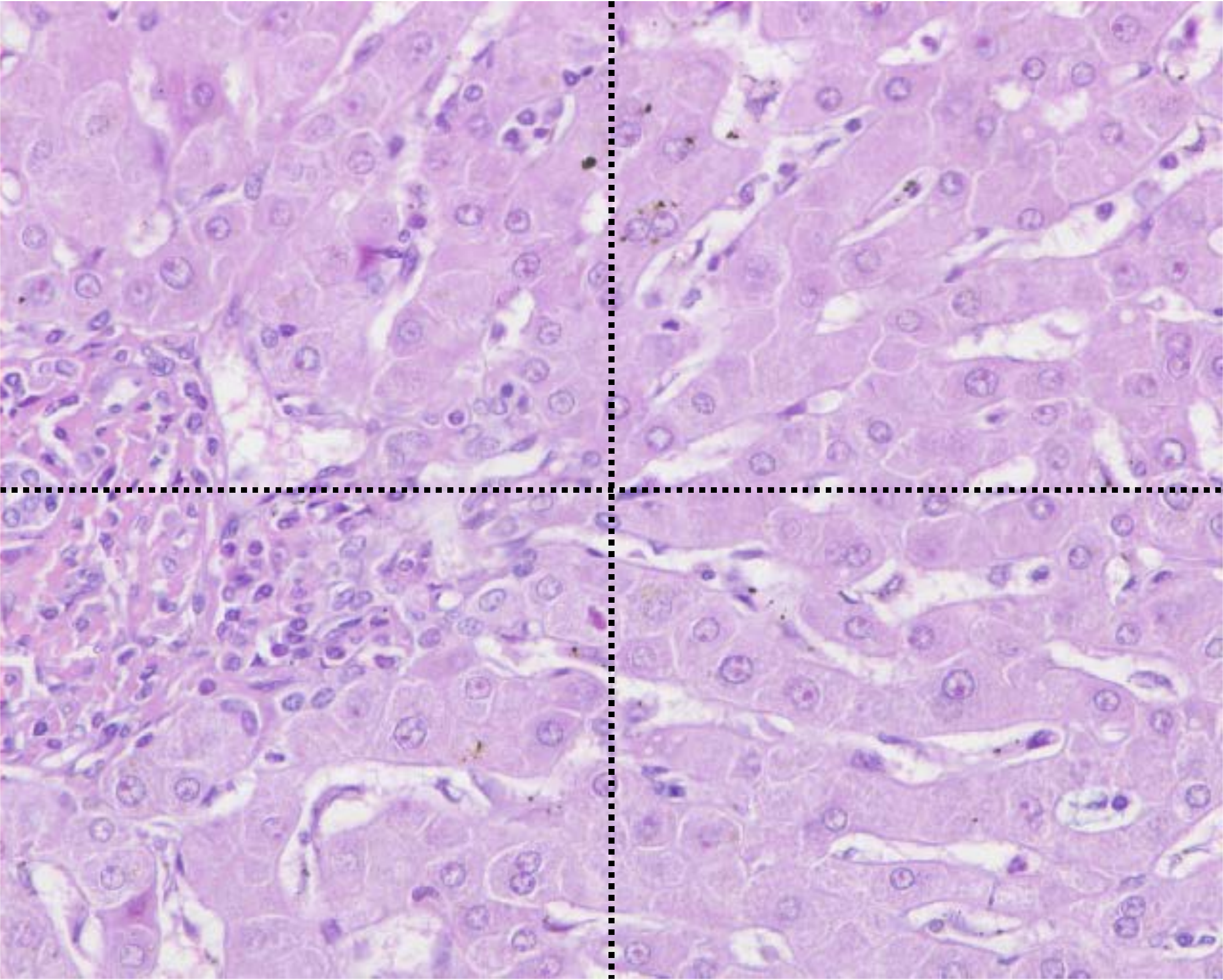}%
  \label{fig:Image1}}
 }
 \centerline{
  \subfloat[]{\includegraphics[width = 0.7\linewidth]{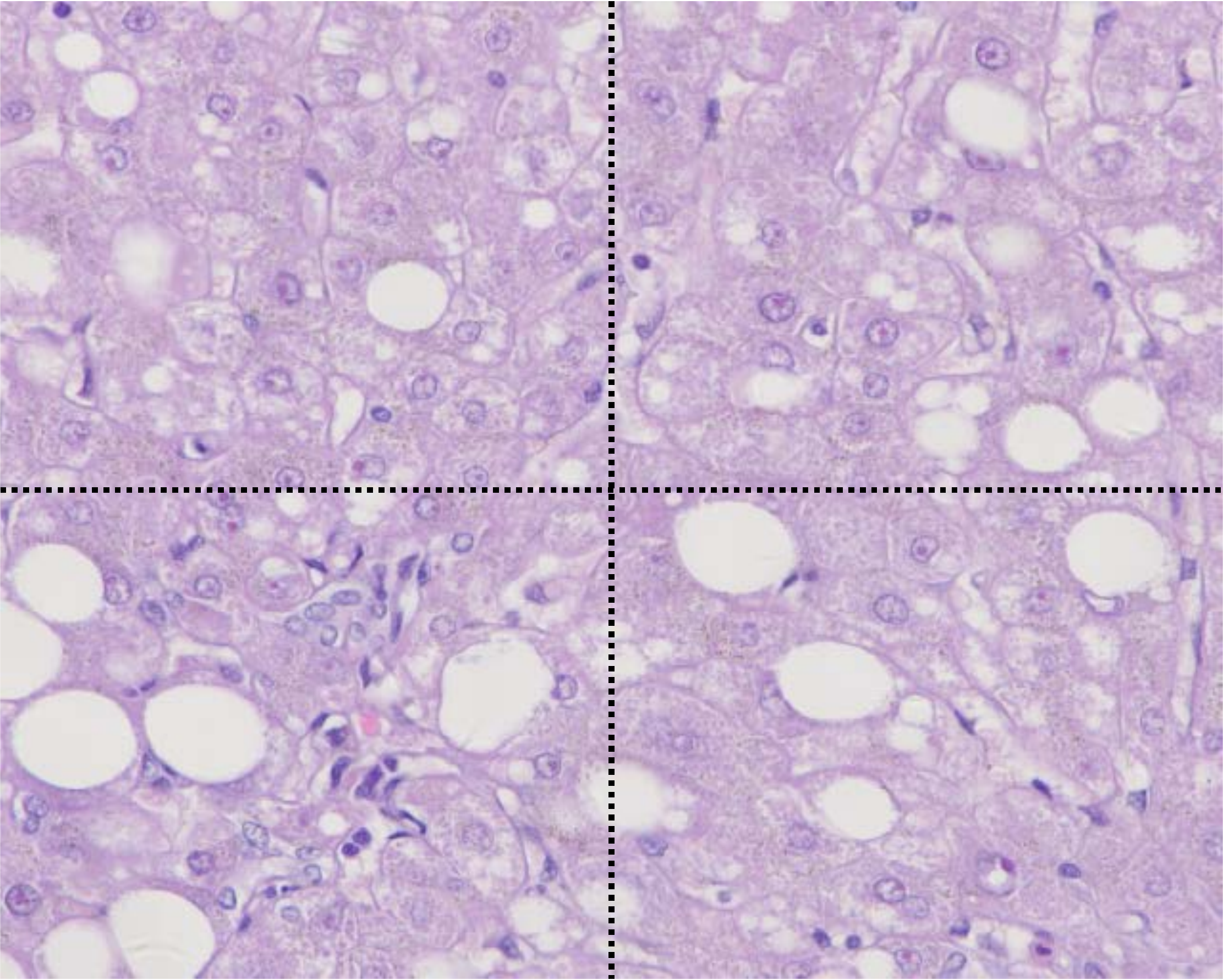}%
  \label{fig:Image2}}
 }
 \centerline{
  \subfloat[]{\includegraphics[width = 0.7\linewidth]{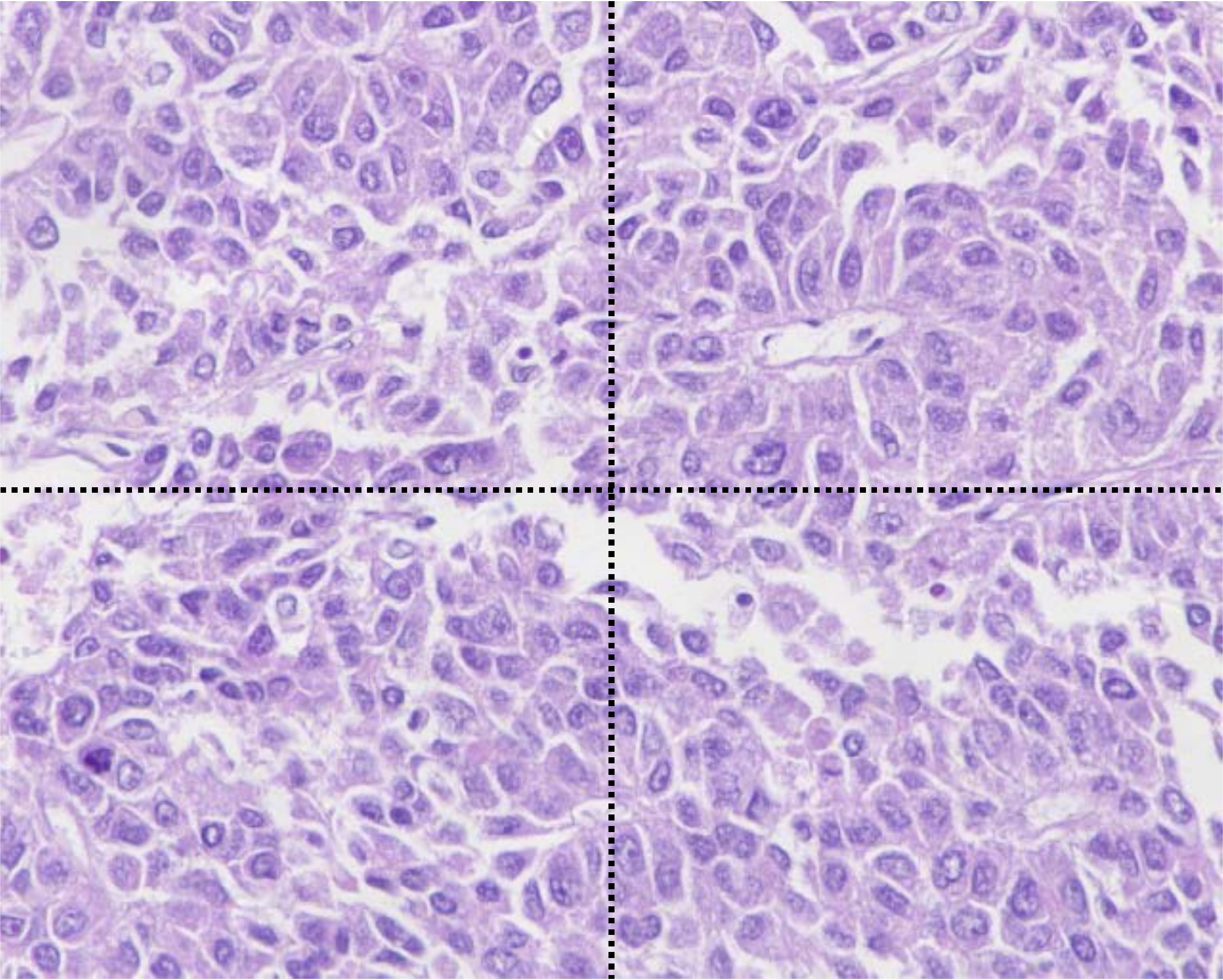}%
  \label{fig:Image3}}
 }
 \caption{Test images (a) Image 1, (b) Image 2, (c) Image 3}
 \label{fig:TestImages}
\end{figure}

\begin{figure}[!t]
 \begin{center}
  \includegraphics[width = 0.65\linewidth]{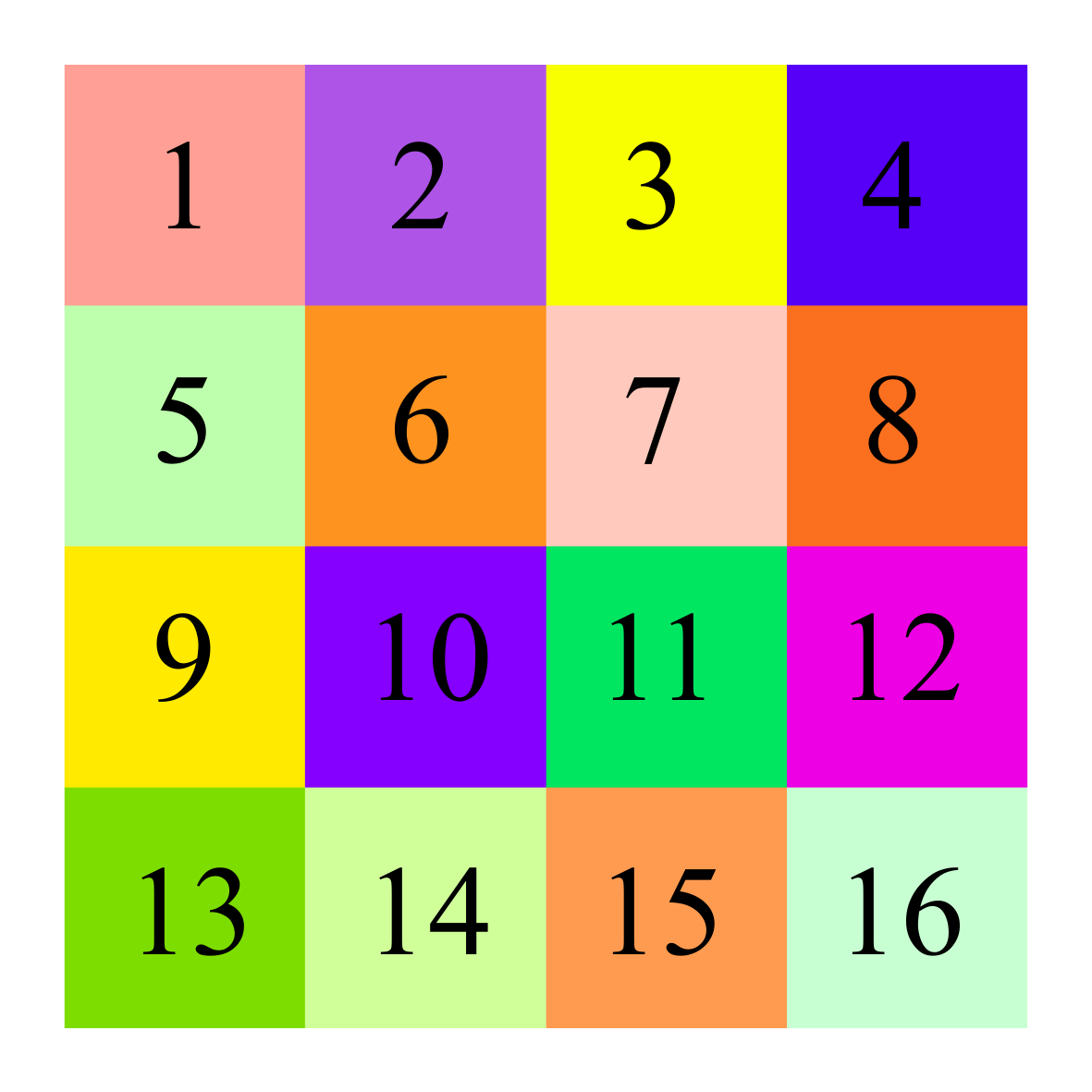}
 \end{center}
 \caption{Optimized MSFA using the proposed method}
 \label{fig:MSFA_proposed}
\end{figure}

\begin{figure*}[!t]
 \centerline{
  \subfloat[]{\includegraphics[width = 0.45\linewidth]{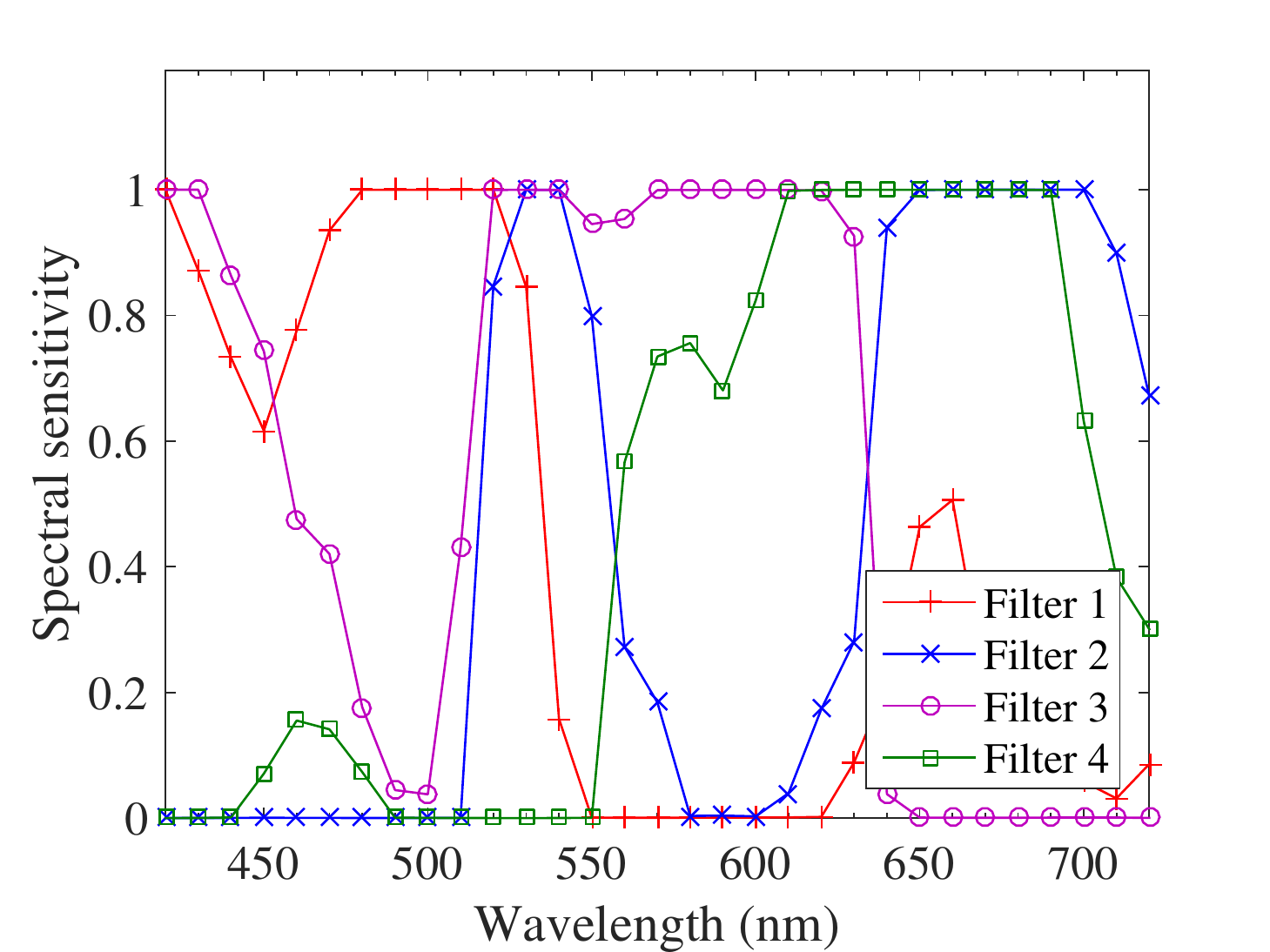}%
  \label{fig:OurFilter1}}
  \hfil
  \subfloat[]{\includegraphics[width = 0.45\linewidth]{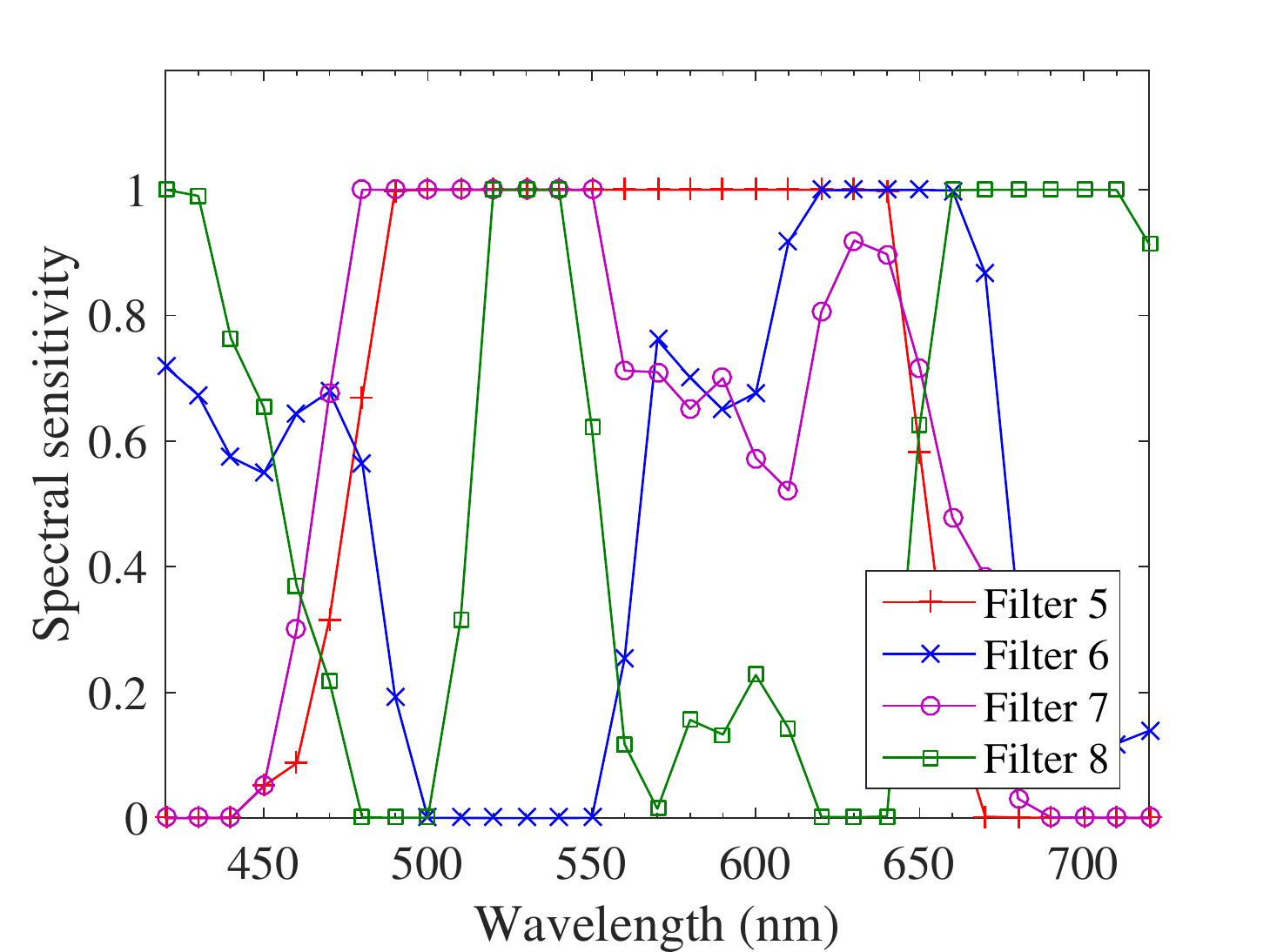}%
  \label{fig:OurFilter2}}
 }
 \centerline{
  \subfloat[]{\includegraphics[width = 0.45\linewidth]{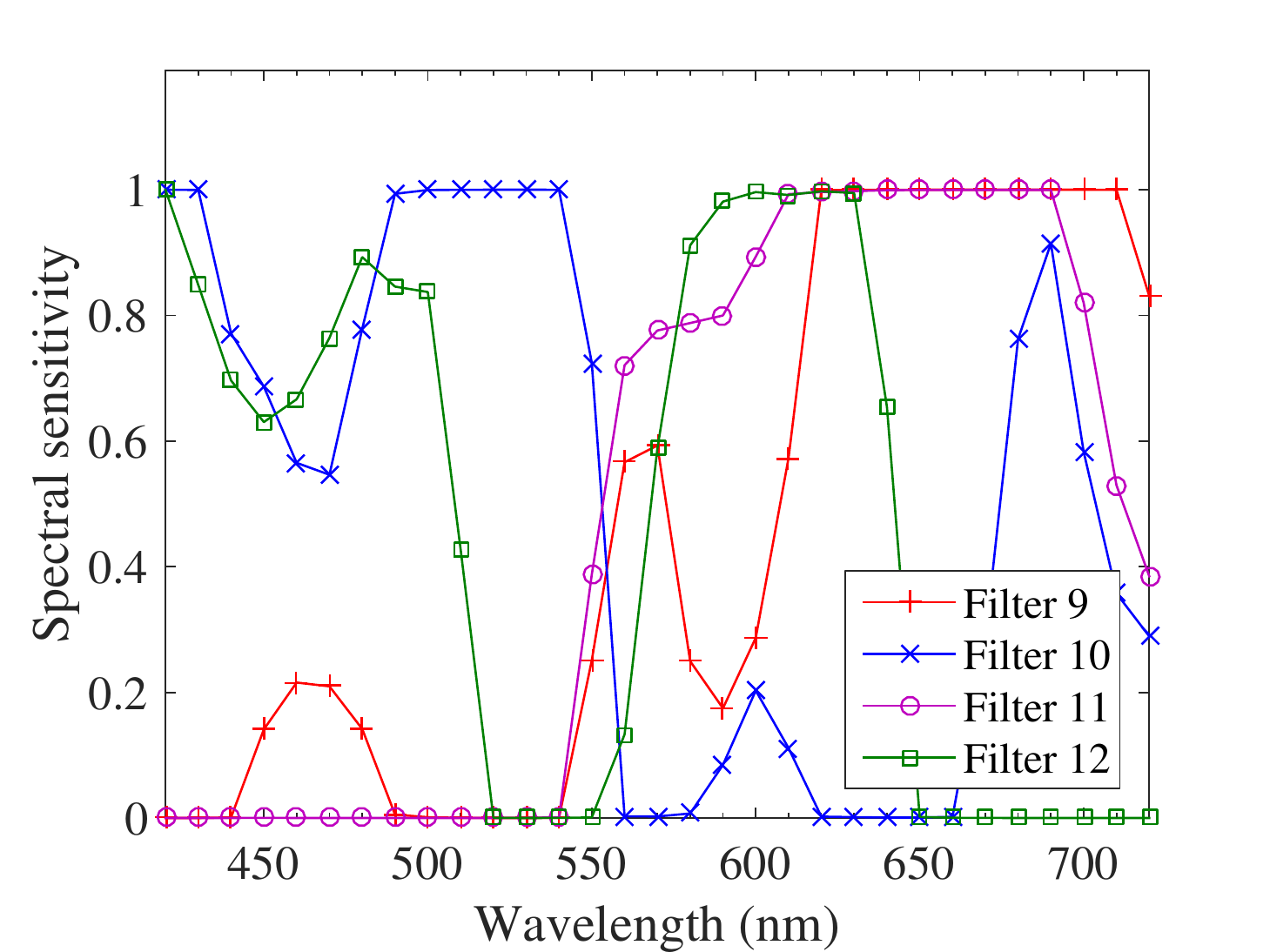}%
  \label{fig:OurFilter3}}
  \hfil
  \subfloat[]{\includegraphics[width = 0.45\linewidth]{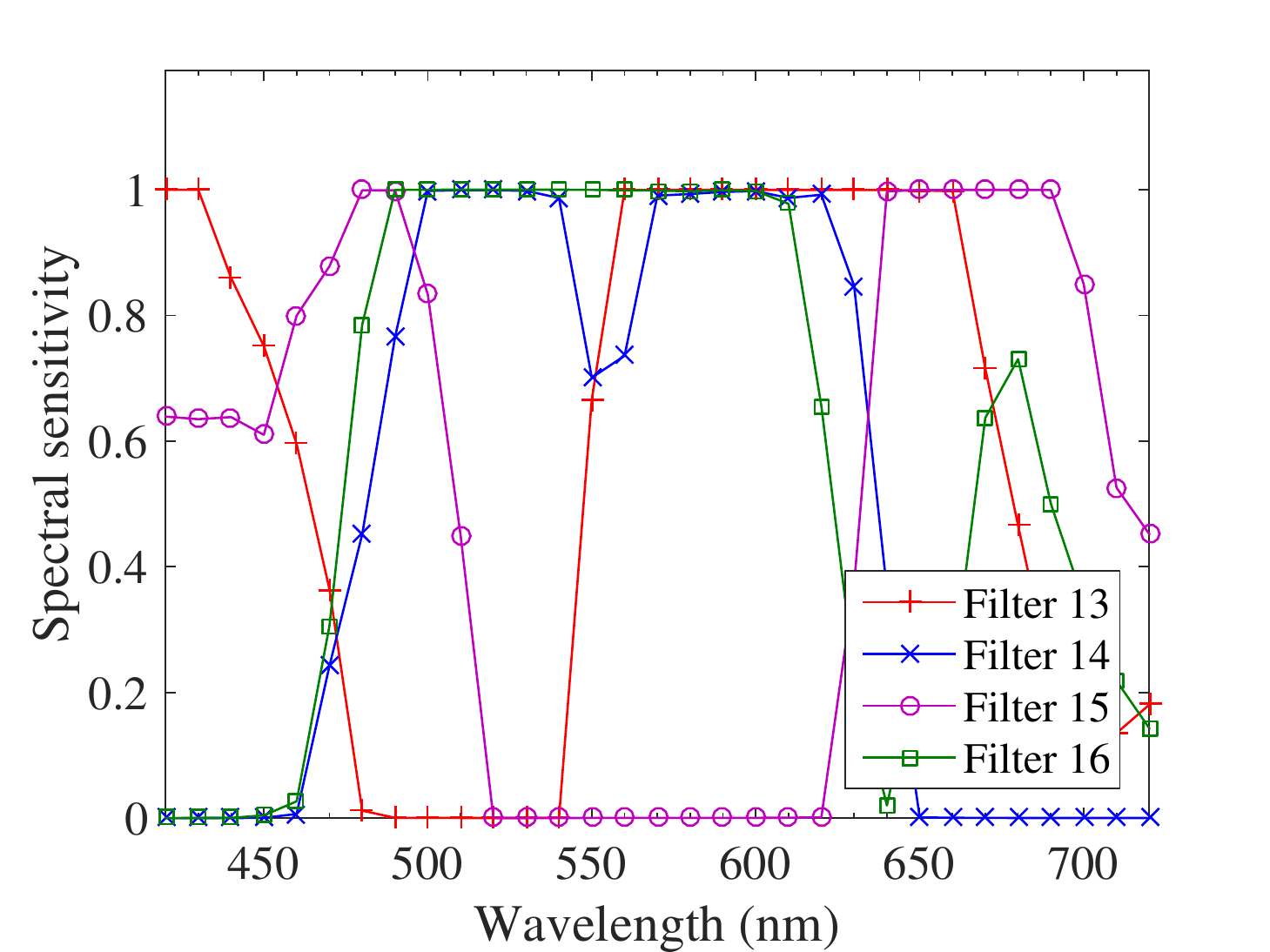}%
  \label{fig:OurFilter4}}
 }
 \caption{Optimized spectral sensitivity of the proposed MSFA (a) filters 1 to 4, (b) filters 5 to 8, (c) filters 9 to 12, and (d) filters 13 to 16}
 \label{fig:SpectralSense}
\end{figure*}

\begin{figure}[!t]
 \centerline{
  \subfloat[]{\includegraphics[width = 0.45\linewidth]{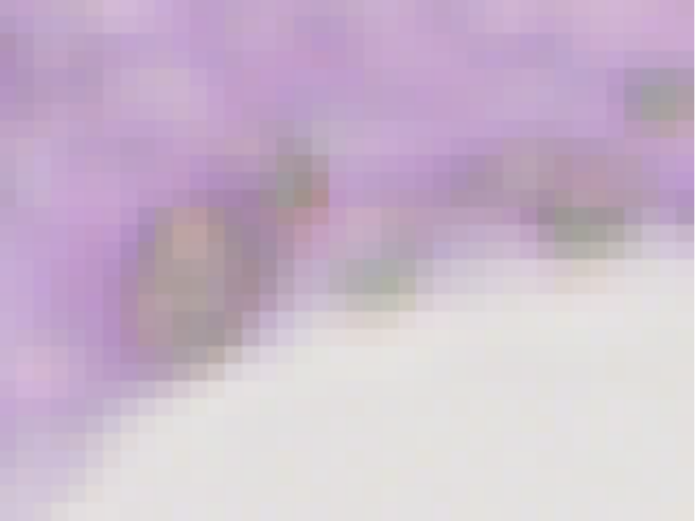}%
  \label{fig:partial02_true}}
  \hfil
  \subfloat[]{\includegraphics[width = 0.45\linewidth]{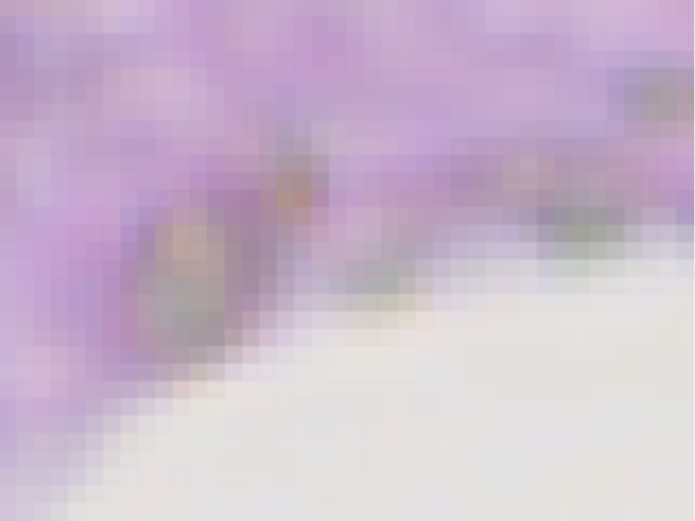}%
  \label{fig:partial02_rgb}}
 }
 \centerline{
  \subfloat[]{\includegraphics[width = 0.45\linewidth]{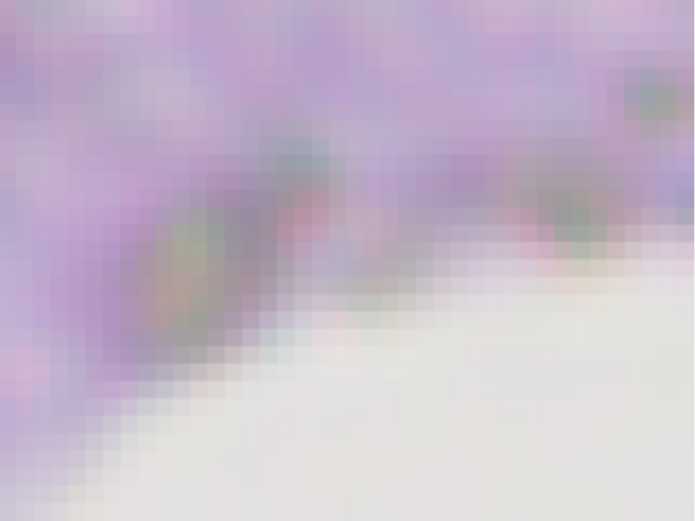}%
  \label{fig:partial02_monno}}
  \hfil
  \subfloat[]{\includegraphics[width = 0.45\linewidth]{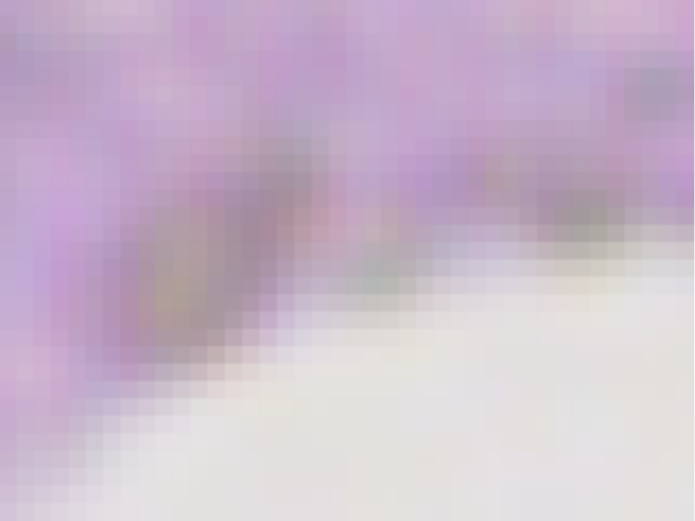}%
  \label{fig:partial02_bay}}
 }
 \centerline{
  \subfloat[]{\includegraphics[width = 0.45\linewidth]{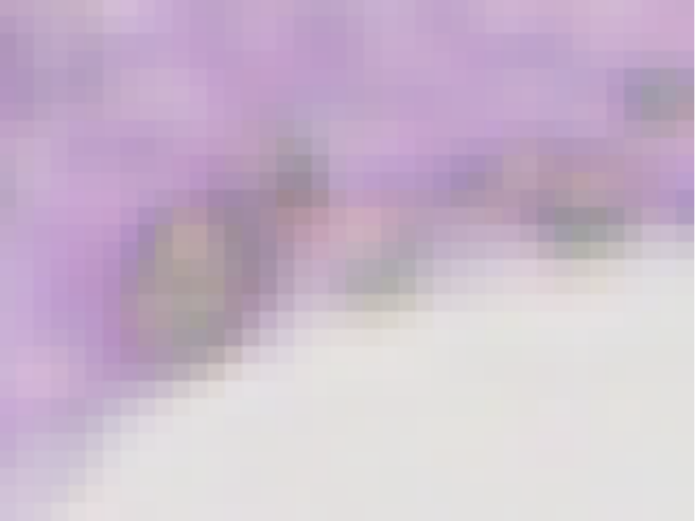}%
  \label{fig:partial02_ours}}
 }
 \caption{Comparison of demosaicked RGB images (Image 2, label 1) (a) original, (b) RGB Bayer, (c) Monno, (d) bandpass, and (e) proposed}
 \label{fig:RGBcomp}
\end{figure}

\begin{figure}[!t]
 \centerline{
  \subfloat[]{\includegraphics[width = 0.9\linewidth]{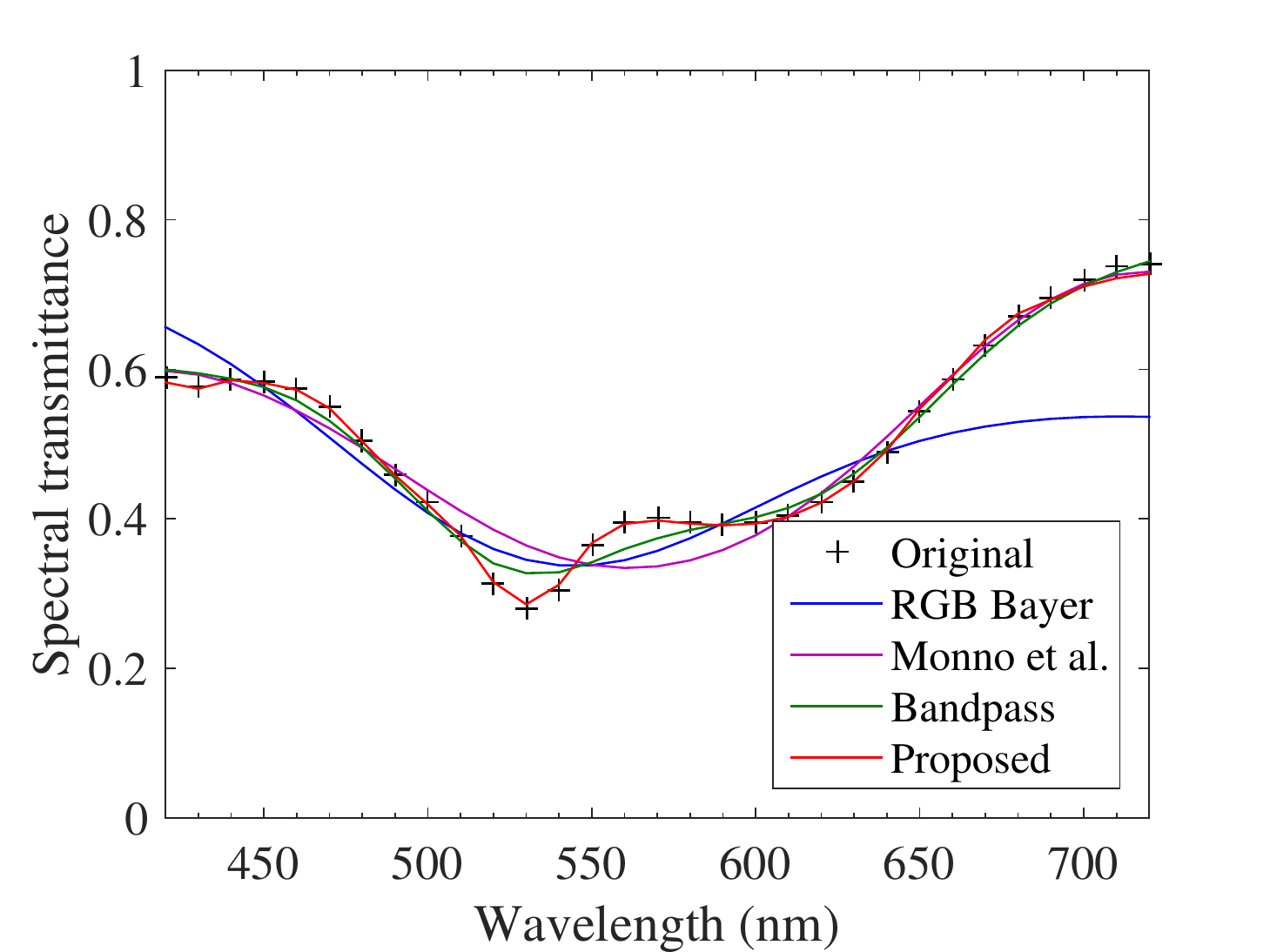}%
  \label{fig:Nuclear}}
 }
 \centerline{
  \subfloat[]{\includegraphics[width = 0.9\linewidth]{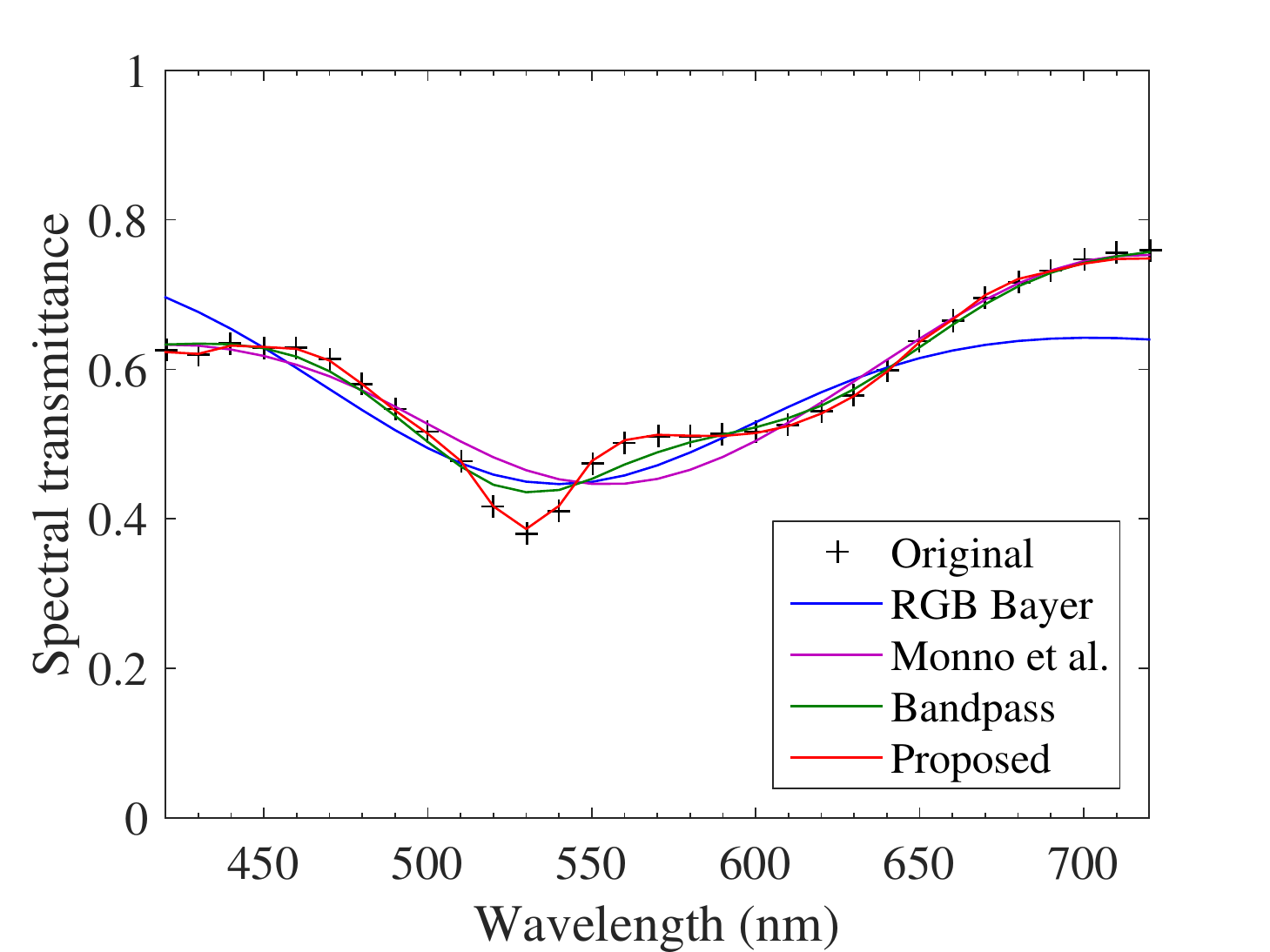}%
  \label{fig:Cytoplasm}}
 }
 \caption{Average spectrum of demosaicked MSI (Image 2, label 1) (a) nuclear and (b) cytoplasm}
 \label{fig:AveSpectrum}
\end{figure}

In the experiment, the performances of the optimized MSFA and demosaicking are compared with those using a conventional method. A real MSI, which is not mosaicked, is used as the original image. Subsequently, mosaicking and demosaicking are simulated on a computer. A pathological tissue, namely, human liver tissue $20\times$ optically zoomed and stained with H\&E is captured using a 151-band hyperspectral camera (EBA Japan, NH-7) and an optical microscope (Olympus, BX53), as shown in \figref{fig:Setup}. The captured test images, which are converted from MSI into sRGB, are shown in \figref{fig:TestImages}. 
The captured image is $1280 \times 1024$ pixels, 151 bands, 350 to 1100 nm at 5 nm. To increase the number of test images in the experiments, these images are divided into four sub-images and labeled as follows: upper-left, 1; upper-right, 2; lower-left, 3; and lower-right, 4.
The number of bands is reduced from 151 to 31 (420 to 720 nm at 10 nm intervals) to remove the noise bands. 

Additionally, to increase the number of test images in the following experiments, these three test images are divided into four sub-images and labeled as follows: upper-left, 1; upper-right, 2; lower-left, 3; and lower-right, 4.
Image 1-1 (upper-left part of Image 1) is used for training, and all other 11 test images are used as the test data.
In the following experiments, the demosaicking evaluation of Image 1-1 corresponds to the evaluation of the learning capabilities, and the evaluation of the other images correspond to the versatility performance of the optimized MSFA.

\subsection{Optimized MSFA pattern and spectral sensitivity}
The optimized filter pattern is shown in \figref{fig:MSFA_proposed}, and the spectral sensitivities are shown in \figref{fig:SpectralSense}. Note that each pixel in \figref{fig:MSFA_proposed} is colored according to the sRGB color reproduction under a D65 illuminant, and the sensitivities correspond to $\Vec{\phi}_n$ ($n$ = 1 to 16) in \eqref{eq:Phi}. Each filter of the optimized MSFA can be seen as a mixture of colors because the spectral sensitivity functions become relatively broad, and the appearance of these filters is slightly different compared with that of the conventional MSFAs in \figref{fig:ExamplesOfMSFA}. Because the optimized spectral sensitivity depends on the training data, it is difficult to predict the ideal sensitivity. It is clear that the optimized spectral sensitivity obtained from a heuristic approach certainly improves the demosaicked image quality, but it is difficult to understand the meaning of the obtained spectral properties. However, it should be noted that Hirakawa et al. \cite{ref:KHirakawa2008} mentioned that a mixture of colors suppresses the aliasing in the case of RGB color filter arrays, and therefore our optimized broadband (mixed wavelengths) filters are deemed suitable for MSFAs as well.

\subsection{Demosaicked image quality in terms of PSNR}

\begin{table*}[t]
\caption{PSNR [dB] comparison of demosaicked MSIs}
 \label{tab:PSNR}
\centering
  \begin{tabular}{c|c|c|c|c|c|c} \hline
Test image & Label & RGB Bayer & Monno & Bandpass & Proposed (1block) & Proposed (9blocks)  \\ \hline\hline
Image 1 & 1 (training) & 24.529 & 25.992 & 28.730 & 32.761 & {\bf 32.949}  \\ 
 & 2 & 24.892 & 26.127 & 28.856 & 32.371 & {\bf 32.567}  \\ 
 & 3 & 24.189 & 26.323 & 28.649 & 32.205 & {\bf 32.487}  \\ 
 & 4 & 24.737 & 26.474 & 28.792 & 31.955 & {\bf 32.252}  \\ \hline
Image 2 & 1 & 24.577 & 28.069 & 29.961 & 32.762 & {\bf 33.292}  \\ 
 & 2 & 24.605 & 28.309 & 30.222 & 32.411 & {\bf 32.990}  \\
 & 3 & 24.172 & 28.469 & 30.065 & 32.480 & {\bf 32.954}  \\
 & 4 & 24.583 & 28.321 & 29.984 & 32.026 & {\bf 32.634}  \\ \hline
Image 3 & 1 & 22.526 & 27.450 & 29.328 & 32.523 & {\bf 32.765}  \\ 
 & 2 & 22.193 & 27.147 & 29.177 & 31.927 & {\bf 32.344}  \\ 
 & 3 & 22.045 & 27.319 & 29.21 & 32.237 & {\bf 32.516}  \\
 & 4 & 22.078 & 27.472 & 29.336 & 31.806 & {\bf 32.200}  \\ \hline
  \end{tabular}
\end{table*}

\begin{table*}[t]
\caption{PSNR [dB] comparison of demosaicked RGB}
 \label{tab:PSNR_RGB}
\centering
  \begin{tabular}{c|c|c|c|c|c|c} \hline
Test image & Label & RGB Bayer & Monno & Bandpass & Proposed (1block) & Proposed (9blocks)  \\ \hline\hline
Image 1 & 1 (training) & 42.038 & 34.007 & 37.402 & 41.522 & {\bf 43.168}  \\
 & 2 & 42.087 & 34.337 & 37.676 & 40.840 & {\bf 42.396}  \\ 
 & 3 & 41.333 & 33.906 & 36.582 & 40.485 & {\bf 42.345}  \\ 
 & 4 & 41.699 & 34.433 & 37.113 & 40.390 & {\bf 42.071}  \\ \hline
Image 2 & 1 & 41.794 & 35.484 & 37.838 & 41.198 & {\bf 43.302}  \\ 
 & 2 & 42.048 & 35.897 & 38.286 & 40.788 & {\bf 42.880}  \\ 
 & 3 & 41.388 & 35.495 & 37.707 & 40.818 & {\bf 42.680}  \\ 
 & 4 & 41.672 & 35.654 & 37.702 & 40.040 & {\bf 42.090}  \\ \hline
Image 3 & 1 & 41.476 & 34.238 & 36.677 & 40.738 & {\bf 42.302}  \\ 
 & 2 & {\bf 41.630} & 33.894 & 36.429 & 39.757 & 41.499  \\
 & 3 & 41.191 & 33.861 & 36.617 & 40.347 & {\bf 41.898}  \\
 & 4 & {\bf 41.442} & 34.136 & 36.749 & 39.779 & 41.383  \\ \hline
  \end{tabular}
\end{table*}

The peak signal-to-noise ratios (PSNRs) of the demosaicked 31-band MSI is shown in \tabref{tab:PSNR}.
Here, the PSNR using the RGB Bayer color filter array, the 6-band MSFA proposed by Monno et al. \cite{ref:YMonno2015}, and the 16-band MSFA using only bandpass filters chosen from 420 to 720 nm at constant 20 nm intervals are compared.
These three methods use Wiener estimation based on a first-order Markov chain \cite{ref:KShinoda2017} as non-trained demosaicking.
``Proposed (1block)'' means the demosaicking performance using only one block for the Wiener estimation based on \eqref{eq:Demosaicking}, whereas ``Proposed (9block)'' uses the center and neighboring eight blocks for the Wiener estimation based on \eqref{eq:Demosaicking3}.
Here, note that the PSNR of Image 1-1 is also shown to confirm the learning ability of the proposed method.
The proposed method using nine blocks achieves approximately a 4 to 6 dB improvement in MSI compared to the Monno MSFA, and a 3 to 4 dB improvement compared to the Bandpass MSFA.
Although the Monno MSFA was designed to improve the spectrum reconstruction, its spectral sensitivity and demosaicking are optimized for natural images, but not for pathological images. The bandpass MSFA exhibits a slightly higher PSNR than the 6-band MSFA because the number of bands is increased; however, its spectral sensitivity and demosaicking are not optimized for pathological images, in contrast to the Monno MSFA.
Therefore, the proposed joint optimization method for MSFA and demosaicking has a considerable advantage in terms of spectrum reconstruction. 

Additionally, there are almost no significant difference between Image 1-1 (training) and the others in \tabref{tab:PSNR}. This implies that the generalization ability of the proposed method is comparable to the learning ability.
Even if the test image is different, the spectrum of the H\&E-stained pathological images has a similar form because the tissues contain only two types of dye.
Therefore, in pathological images, the proposed method is considerably more versatile with a limited number of training images.

The PSNR of the demosaicked RGB image is shown in \tabref{tab:PSNR_RGB}.
The PSNR of the RGB image in the proposed method is also higher than that in the other methods, as well as the MSI.
It is remarkable that the proposed PSNR is slightly higher than that of the RGB Bayer in Images 1 and 2.
For the reproduction of an sRGB image, RGB Bayer has an advantage because its spectral sensitivity consists of only R, G, and B. In contrast, the spectral sensitivity of the proposed MSFA does not contain the RGB components only. Because the proposed method can recover a more precise spectrum than the other methods, the RGB reproduction is also improved.

\subsection{Visual comparison of demosaicked image}
A demosaicked RGB image is shown in \figref{fig:RGBcomp}.
Although the images achieved using the Monno and Bandpass MSFAs differ from the others, there are nearly no visual differences in the images between RGB Bayer and the proposed method.
However, a performance difference can be seen more clearly in the comparison of the average spectrum, as shown in \figref{fig:AveSpectrum}.
Here, the averaged samples are chosen from the demosaicked MSI of Image 2-1.
RGB Bayer can reproduce a precise sRGB image; however, the recovered spectrum has large errors at all wavelengths.
The error of the Bandpass MSFA becomes smaller than that of the Monno MSFA before and after 550 nm; however, the valley cannot be recovered at 530 nm.
It is clear that the proposed method reproduces the original spectrum in both the nuclear and cytoplasm regions.
Consequently, it can be used for a pathology-oriented MSFA, and the demosaicked MSI, RGB, and spectrum quality can be significantly improved.

\subsection{Computational complexity}

\begin{figure}[!t]
 \begin{center}
  \includegraphics[width = 0.9\linewidth]{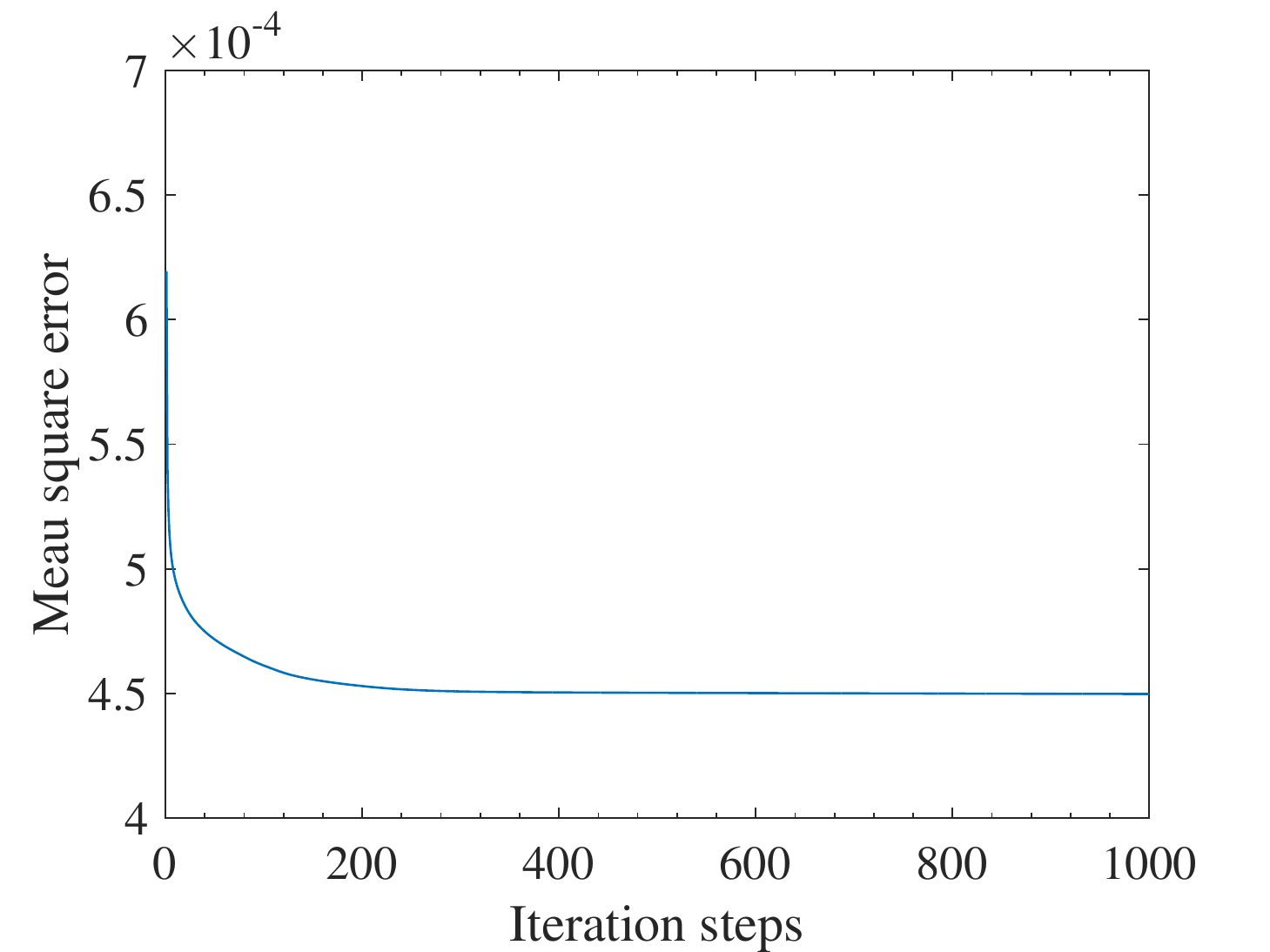}
 \end{center}
 \caption{Convergence graph of mean square error with the proposed method}
 \label{fig:Convergence}
\end{figure}

A convergence graph of the mean square error of the proposed method is shown in \figref{fig:Convergence}. It is clear that the proposed iteration algorithm can effectively reduce the error, and is sufficiently converged within 1,000 iterations. With the proposed method, the processing time for 1,000 iterations is 1,464 s (Intel core i7-3770, 3.40 GHz). Most of the time was consumed within the interior-point approach of the {\it fmincon} function of Matlab, and amounts to about 832 s; the remaining time was consumed a data-type conversion such as {\it mat2cell} function.

The computational time for designing the MSFA is irrelevant in practice because the design process is executed only once, namely, at the initial stage of production. However, reducing the processing time can result in a more efficient approximation of the global minimum. This improvement in the optimization algorithm, and implementing a parallel computing algorithm, are aspects of our future work.

\section{Conclusion}
A new method for optimizing the spectral sensitivity and demosaicking of MSFAs was proposed, and used H\&E-stained pathological images as training data.
The MSFA pattern and demosaicking matrix can be optimized by minimizing the error between the original and demosaicked images during a simulation, which involves formulating an MSFA as an optimization problem and providing an iterative procedure to locally obtain the optimal solutions.
During the experiment, the optimized MSFA pattern and the demosaicked MSI were compared with the conventional MSFA, and it was demonstrated that they achieve a higher PSNR for both MSIs and RGB images.
This implies that the proposed imaging method can greatly contribute to the development of an MSI-based medical imaging application because the proposed MSFA and demosaicking are advantageous not only to MSIs but also to RGB images.
The proposed imaging system can capture both MSIs and RGB images (RGB images can be obtained from MSIs), its image quality is higher than that of a conventional MSFA and an RGB Bayer filter array, and the size of the camera equipped with an MSFA is nearly the same as that of current RGB cameras because only the filter array is replaced.

Because an H\&E stain is one of the most widely used stains in medical diagnoses, an optimized MSFA has the potential to be applied to many kinds of tissues, not only human liver.
Additionally, although the optimized MSFA of this study was specialized using H\&E-stained pathological tissue, the proposed method can also be applied to other stains by changing the training data.
In future work, considerably larger numbers of training and test images will be used for evaluating the generalization ability, and the manufacturing costs of highly complicated and sensitive MSFAs will be considered.

For the applied algorithm, although we used $3 \times 3$ block-based demosaicking with Wiener estimation, the demosaicking algorithm and the size of the blocks used should also be improved.
We will also focus on a proximal gradient method for solving the quadratic programming problem. 

\section*{Acknowledgment}
This work was supported by JSPS KAKENHI Grant Nos. JP26108002 and 15K20899, the Telecommunications Advancement Foundation, Grant for Advancement of Scientific Research and Development, and Konica Minolta Imaging Science Encouragement Award.

\ifCLASSOPTIONcaptionsoff
  \newpage
\fi

\end{document}